\documentclass[a4paper,12pt]{amsart}
\usepackage{natbib}
\usepackage{bibunits}
\usepackage{amsmath}
\usepackage[obeyspaces]{url}
\usepackage{epigraph}

\title{Quantum profiles and paradoxes}
\thanks{Paper originally published in Swedish as ''Kvantumprofiler och paradoxer'', \textit{Arkhimedes} (Helsinki), no. 3, 1992:166-189. Present address of the author: University of Jyv\"askyl\"a, Chydenius Institute, POB 567, 67101-Karleby, Finland; email: \url{borgbros@netti.fi}.}
\author{Frank Borg}

\begin{document}
\begin{bibunit}[apalike]

\begin{abstract} This paper discusses questions concerning the foundations of quantum mechanics (entanglement, wave collapse, irreversibility) with reference to the issues raised during a Minisymposium held in Helsinki, 1.6-3.6 in 1992, where A Shimony, A Peres and B d'Espagnat were invited to lecture. The measurement problem is related to the phenomenon of irreversibility which is known not to follow from any fundamental theory; e.g., the law of exponential decay does not follow from the Schr\"odinger equation in senso stricto. The approximations that lead to irreversibility are related to some form of ''forgetting'', or coarse graining. Some approaches to the question, why these approximations can be justified, are reviewed. The paper also discusses dynamical reduction schemes and it is suggested, that the corresponding non-linear modifications of the Schr\"odinger equation -- which lead to non-conservation of energy -- are actually ''effective'' Schr\"odinger equations for open systems (in analogy with the ''effective'' Lagrangians in QFT). Thus these modifications should be derived from a fundamental theory of interactions. The paper ends with some brief comments on the mind-body question in the quantum mechanical context. A postscript has been added (2006).

\end{abstract}

\maketitle

\tableofcontents

\epigraph{The problem of the meaning of ''reality'' is too much philosophical to be addressed and resolved by the present paper}{\citet{Namiki1991}}
\section{Introduction}

In his classic text \citep{Jammer1974} on the various interpretations of quantum mechanics Jammer comments -- after having devoured some 500 authors and their proposed solutions to the quantum enigmas:

\begin{quotation}
The endless stream of publications suggesting new theories of measurement and modifications of ideas earlier proposed as well as the unending discussions and symposia on this topic are an eloquent indication of the general discomfort felt among the physicists on this issue.	
\end{quotation}

Finland, a country of a thousand conferences, has also assumed its responsibility for keeping the quantum symposia carousel spinning. The city of Joensuu has been an active part sponsoring a series of conferences on \textit{The Foundations of Modern Physics} (1977, 1985, 1987, 1990). However, this year even Joensuu is hemorrhaging because of the general economic backlash. Despite this it was possible to arrange a minispymposium during some sunny days in the beginning of June at the physics department of the university of Helsinki with its kind support \citep{Laurikainen1993}. The ''stars'' of the symposium consisted of the triumvirate made up of Bernard d'Espagnat (Orsay, France), Abner Shimony (Boston, USA) and Asher Peres (Haifa, Israel). d'Espagnat has achieved a reputation (vide \cite{Espagnat1984,Espagnat1988,Espagnat1989}) as an incisive analyzer of the conceptual-philosophical foundations of quantum mechanics, and for having introduced the concept of the ''veiled reality'' which refers to something that cannot by studied by traditional scientific methods -- he suggests that poetry maybe offers a link to this reality. Peres started his physics career as nuclear engineer. He gives the impression of a hardnosed problem solver but with talents of a comedian not unlike Woody Allen. Other physicists have classified him as a positivist, among them Abner Shimony who is a very jovial professor who feels at home among neutrons. He describes himself as student of Carnap, and he is able to give a penetrating lecture on some subtle points by Kant or Berkeley at any time. Car engines he finds, however, to be difficult to understand; indeed, this involves an important philosophical point (complicated phenomena should be reduced to simple constituents which are easier to understand). With the lectures of the symposium as a background we will in the next sections discuss some central issues in quantum mechanics. 

\epigraph{I like the concept of the World-Soul, because it is deeply poetical}{d'Espagnat, at the conference}
\section{Open realism}

d'Espagnat defines his philosophical view as ''open realism''; existence precedes knowledge; something exists independently of us even if it cannot be described. Berkeleyean idealism, phenomenalism, positivism, anti-realism, internal realism and similar philosophies assume that one only observes mental pictures, not objects, whence there is no need to presuppose the reality of objects per se. Shimony summarizes Berkeley's arguments for idealism in two points:

\begin{itemize}
	\item[a] The description of things only involves thoughts 
	\item[b] Matter (quantity) cannot explain perception (qualia)
\end{itemize}
 
d'Espagnat's objections against idealism rest on two propositions:

\begin{itemize}
	\item[a] Experience and knowledge implicate the existence of consciousness (mind) 
	\item[b] One may be wrong (reality ''says no''); that is, just any theories do not work, there is something (''objective reality'') which resists
\end{itemize}

Coming as far as Samuel Johnson (kicking the famous stone to refute Berkeley) and accepting a reality, d'Espagnat asks whether this reality is intelligible. Indeed, quantum mechanics challenges any physicist-philosopher with realist inclinations, which Albert Einstein realized from early on. Einstein pointed to the apparently instantaneous ''telepathic'' collapse of the wave functions; an aspect that was brought to its head in the famous Einstein-Podolsky-Rosen paper \citep{EPR1935}. \citet{Schroedinger1935} concluded immediately that the fundamental point of the EPR-argument was the ''entanglement'' of the composite system. Thus, two systems $S_1$ and $S_2$ constitute through interaction a total system $S_1 + S_2$ whose state vector in general is of the form

\begin{equation}
\label{EQ:1} 
\left| \Psi \right\rangle = \sum c_{nm} \left| \phi_n \right\rangle \otimes \left| \eta_m \right\rangle
\end{equation}

where $\left| \phi_n \right\rangle$ ($\left| \eta_m \right\rangle$) belong to the Hilbert space $H_1$ ($H_2$) for the system $S_1$ ($S_2$). If we, as Schr\"odinger did, identify the state of an object/system with its statevector, it follows that for an entanglement state (\ref{EQ:1}) (which in general cannot be factorized on the form $\left| \phi \right\rangle \otimes \left| \eta \right\rangle$) the subsystems $S_1$ and $S_2$ will lack definitive physical states. This conclusion also applies with regards to the physical properties if we identify these with the eigenvalues of the corresponding operators. Therefore, even if the total system $S_1 + S_2$ is in a definitive state this does not in general apply for its subsystems. This consequence makes quantum mechanics, according to d'Espagnat quoting H Weyl, ''the first holistic theory that works''. At the same time this property creates problems for every attempt to make an ontological interpretation of quantum mechanics which ascribes to all systems/objects definitive physical states and properties at every instant of time.

The entanglement situation arises also in the case where the system $S_1$ is an object which interacts with an apparatus $S_2$ during a measurement process. In order that the process should result in a definitive state $\left| \eta \right\rangle$ of the measurement apparatus it is necessary that the superposition Eq.(\ref{EQ:1}) is reduced on the factorized form
$\left| \phi \right\rangle \otimes \left| \eta \right\rangle$ which is generally impossible in quantum mechanics. Indeed, the quantum mechanical evolution $\left| \Psi \right\rangle \rightarrow \left| \Psi \right\rangle_t = U_t \left| \Psi \right\rangle$ is linear and preserves superpositions,
$U_t (\left| \Psi_1 \right\rangle + \left| \Psi_2 \right\rangle) = U_t \left| \Psi_1 \right\rangle + U_t \left| \Psi_2 \right\rangle$. Assume that an ideal measurement process transforms the initial state $\left| \phi_i \right\rangle \otimes \left| \eta_0 \right\rangle)$ into $\left| \phi_i \right\rangle \otimes \left| \eta_i \right\rangle)$:

\begin{equation}
\label{EQ:2}
\left| \phi_i \right\rangle \otimes \left| \eta_0 \right\rangle \rightarrow U_t (\left| \phi_i \right\rangle \otimes \left| \eta_0 \right\rangle) = \left| \phi_i \right\rangle \otimes \left| \eta_i \right\rangle.
\end{equation}
 
Here $\left| \eta_0 \right\rangle$ denotes the initial state of the measurement apparatus while its state $\left| \eta_i \right\rangle$ registers that the object was in the state $\left| \phi_i \right\rangle$. If we apply the linear property of the quantum evolution on the initial state $ (\sum c_i \left| \phi_i \right\rangle) \otimes \left| \eta_0 \right\rangle$ it follows that the object + apparatus evolves into an entanglement state:  

\begin{equation}
\label{EQ:3}
\left(\sum c_i \left| \phi_i \right\rangle \right) \otimes \left| \eta_0 \right\rangle \rightarrow U_t \left(\left(\sum c_i \left| \phi_i \right\rangle\right) \otimes \left| \eta_0 \right\rangle \right) = \sum c_i \left| \phi_i \right\rangle \otimes \left| \eta_i \right\rangle),
\end{equation}

which contradicts the fact that the measurement apparatus has to be in a definitive state. This entangled situation is the central issue of the problem of measurement in quantum mechanics.

d'Espagnat distinguishes three genuinely ontological approaches to quantum mechanics:

\begin{enumerate}
	\item Pilot-wave theories (de Broglie, D Bohm)
	\item Other hidden variable theories \citep{Belinfante1973}
	\item Theories of dynamical reduction \citep{Ghirardi1986,Pearle1984,Karolyhazy1986} 
\end{enumerate}

According to the pilot-wave theories \citep{Bohm1952a,Bohm1952b} the objectively existing particles are directed by a field $\Psi$ which, at least approximately, satisfies the Schr\"odinger equation (SE). If we write the field $\Psi$ as

\begin{equation}
\label{EQ:4}
\Psi = R e^{iS/\hbar},
\end{equation} 

with real functions $R$ and $S$, then the SE is decomposed into two equations:

\begin{eqnarray}
\label{EQ:5}
\frac{(\nabla S)^2}{2m} + V - \frac{\hbar^2}{2m} \frac{\nabla^2 R}{R} + \frac{\partial S}{\partial t} = 0\\
\nonumber
\nabla \cdot \left( \frac{\nabla S}{m} R^2 \right) + \frac{\partial R^2}{\partial t} = 0.
\end{eqnarray} 

The first line in Eq.(\ref{EQ:5}) corresponds to the classical Hamilton-Jacobi equation with an additional ''quantum potential'' $Q = - \hbar^2 /2m \cdot \nabla^2 R/R$, 

\begin{equation}
\label{EQ:6}
H(\mathbf{p}, \mathbf{x}, t) = \frac{\mathbf{p}^2}{2m} + V(\mathbf{x},t) - \frac{\hbar^2}{2m} \frac{\nabla^2 R(\mathbf{x},t)}{R(\mathbf{x},t)},
\end{equation}

while the second line in Eq.(\ref{EQ:5}) corresponds to a continuity equation. The field $\Psi$ directs the particle according to the following generalization of Newton's equation:\footnote{The Bohmian formulation seems especially suited for describing the quantum-classical transition. J. S. Bell has commended it for its pedagogical merits (see ''Introductory remarks'', \textit{Phys. Rep.} 137 (1) 1986:7-9.}

\begin{equation}
\label{EQ:7}
m \frac{d^2 \mathbf{x}}{dt^2} = \frac{d\mathbf{p}}{dt} = \frac{\nabla S}{dt} = - \nabla(V + Q).
\end{equation}

The equality $\mathbf{p} = \nabla S$ is, according to \citet{Bohm1989}, satisfied only as an average of a statistical process whose probability density during normal circumstance coincides with $R^2 = |\Psi|^2$.

L de Broglie proposed that the wave equation has two sorts of solutions (double-solution theory): the ordinary SE solution which only describes the statistical behaviour of the particles, and a singular solution which represents the localized particles themselves. A O Barut has recently presented a similar theory where the singular solutions of de Broglie corresponds to a soliton, ''small psi'', while the ''big psi'' is the conventional Schr\"odinger wave function which can only be interpreted statistically \citep{Barut1990,Barut1991}.\footnote{Theories that represent particles via wave fields are still ridden by the entanglement problems. If the electron ''current'' is given by $j^{\mu} = e \overline{\Psi} \gamma^{\mu} \Psi$, where $\Psi$ is the electron field, then its current is undefined in entanglement situations since we cannot ascribe a definitive field to the electron. Barut has given an interesting response in a letter to be commented elsewhere.}
Bohm's theory does not say anything about the structure of the particles (except that they have classical properties like momentum and position). His primary objective has been to demonstrate the very possibility of an ontological quantum theory \citep{Bohm1987}. To the group (b) above one might count e.g. theories of the sort presented by Dennis Dieks, Simon Kochen and Richard Healey \citep{Dieks1989,Kochen1985,Healey1989}. Similarly to the group (a) -theories, these theories distinguish between two kinds of states: on the hand the quantum state (represented by the wave function), on the other hand the physical/dynamical state which is always defined for all objects in contrast to the quantum states which may not be defined in entanglement situations. In these approaches the starting point is a special bi-orthonormal decomposition (which goes back to E Schmidt 1907; vide \cite{Neumann1932}) of the entanglement state Eq.(\ref{EQ:1}):

\begin{equation}
\label{EQ:8}
\left| \Psi \right\rangle = \sum d_s \left| \overline{\phi}_s \right\rangle \otimes \left| \overline{\eta}_s \right\rangle
\end{equation}

where $\left| \overline{\phi}_s \right\rangle$ and $\left| \overline{\eta}_s \right\rangle$ are orthonormal basis vectors. In case $|d_s| \neq |d_l|$ ($s \neq l$) this decomposition is unique (the other cases cause trouble). According to Dieks Eq.(\ref{EQ:8}) means that the systems $S_1$ and $S_2$ can be ascribed physical states $\left| \overline{\phi}_s \right\rangle$ and $\left| \overline{\eta}_s \right\rangle$ with the probability $|d_s|^2$. Especially in the case of Eq.(\ref{EQ:3}) this assumption by Dieks implies that the measurement apparatus exists in definitive physical states represented by the eigenstates whose statistical weights are determined by the coefficients $c_i$. Healey suggests the possibility of developing a stochastic-dynamic theory which will describe the evolution of these dynamical/physical states.

The fact that d'Espagnat includes the third group above among the genuine ontological theories may seem a bit surprising. The dynamical reduction theories mainly show that the superposition states for macroscopic systems get suppressed. The entanglement issue remains alive for the micro-objects, as well as also for the macro-objects during the time till the superposition has been reduced beyond practical verification. No ontological states have been ascribed to the subsystems in the entanglement state. The same objection can be raised against theories which explain the reduction in terms of influence of the environment, which d'Espagnat indeed rejects as non-ontological. The difference seems to be that d'Espagnat assumes that the wave function in case of the dynamical reduction theories describes an independent reality (again causing trouble in the entanglement situation), while the environmental theories describe a ''subjective reality''. Yet, both kinds of theories seem equally ambiguous on this point. d'Espagnat however reasons that the environment theories presuppose an ensemble interpretation of quantum mechanics which implies that quantum mechanics cannot describe individual systems and hence must be incomplete; ergo, the environment theories can only explain the wave collapse by extending quantum mechanics with ''hidden variables''. This is the solution chosen e.g. by Dieks who combines his ontological commitment with environment theories (decoherence) in order to explain how the physical states are realized in the measurement process.

For d'Espagnat the wave function is not an element of the physical reality, rather it represents an ensemble. Reality per se (Being) is either ''pure X'' (Kant) or ''veiled''. The former alternative leads to idealism which is rejected by d'Espagnat. Idealism e.g. presupposes that the concepts are evident (Anschaulichkeit, Kant; lumen naturale, Descartes) which seems to be in conflict with the fact that we now and then have to revise our concepts and theories as a consequence of new findings and theoretical developments. Concerning the veiled reality d'Espagnat is only able to tell that it transcends human reason (and is in this sense ''irrational''). It cannot be studied using scientific methods which are restricted to the ''empirical reality''.

\epigraph{I am more at home with the neutron than the motor of my car}{A Shimony, at the conference} 
\section{Dynamic reductions}

Professor Shimony plans to devote the next few years attempting to derive the wave collapse by manipulating the SE. The Ghirardi-Weber-Rimini (GWR) ansatz constitutes the archetypical model \citep{Ghirardi1986}. The non-linear GWR-modification of the SE can be interpreted as a stochastic process where the wave function for each particle is reduced, according to a Poisson-process with a characteristic time constant $1/\lambda \approx 10^{16}$ s, on a wave function which localizes the position to a region of the dimension $a \approx 10^{-7}$ m (vide \cite{Bell1987}). That is, the wave function $\Psi(\mathbf{r},t)$ evolves according to the SE except for random occasions (separated by an average time $1/\lambda$) where it makes a ''quantum jump'' and is reduced on the form

\begin{equation}
\label{EQ:9}
\acute{\Psi}(\mathbf{r},t) = \frac{e^{{-|\mathbf{r}-\mathbf{\acute{r}}|^2}/2a^2} \cdot \Psi(\mathbf{r},t)}{\sqrt{\int e^{{-|\mathbf{s}-\mathbf{\acute{r}}|^2}/a^2} \cdot \left|\Psi(\mathbf{s},t)\right|^2 d\mathbf{s}}},
\end{equation}

according to the GWR-ansatz. The reduction center $\mathbf{\acute{r}}$ is chosen randomly according to the probability density

\begin{equation}
\label{EQ:10}
p_t(\mathbf{\acute{r}}) = \frac{1}{a^3 \pi^{3/2}} \int e^{{-|\mathbf{s}-\mathbf{\acute{r}}|^2}/a^2} \cdot \left|\Psi(\mathbf{s},t)\right|^2 d\mathbf{s}. 
\end{equation}

In terms of the density matrix $\rho$ the GWR-process implies that during an average period of $1/\lambda$ the $\rho$ is replaced by $T(\rho)$ with the Gaussian weighing Eq.(\ref{EQ:10}); that is, $T(\rho) \approx \rho(t + 1/\lambda) \approx 1/\lambda \cdot d\rho/dt + \rho(t)$ yielding $d\rho/dt \approx \lambda(T(\rho) - \rho)$. Taking also into account the ordinary SE, $d\rho/dt = - i/\hbar \cdot [\hat{H}, \rho]$, we obtain the complete equation,

\begin{equation}
\label{EQ:11}
\frac{d\rho}{dt} = - \frac{i}{\hbar} \left[ \hat{H}, \rho \right] + \lambda \left(T(\rho) - \rho \right).
\end{equation}

(Nota bene -- by using density matrices we are restricted to describing the statistical development. In the discussions of the GWR-models it is not always clear how they interpret the wave function. In case it represents the physical state we are still left with the entanglement problem for micro-objects. The ensemble interpretation seems to be the other alternative but this interpretation denounces the collapse problem from the very outset -- vide \citet[ch.10]{Jammer1974} and \citet{Home1992}.) The Gaussian weighing is defined by

\begin{equation}
\label{EQ:12}
T(\rho) = \frac{1}{a^3 \pi^{3/2}} \int e^{{-|\mathbf{\hat{q}}-\mathbf{\acute{r}}|^2}/2a^2} \rho(t) e^{{-|\mathbf{\hat{q}}-\mathbf{\acute{r}}|^2}/2a^2} d\mathbf{r},
\end{equation}
   
where $\mathbf{\hat{q}}$ is the position operator. The GWR-process has no observable consequences for an elementary particle since the reduction happens only about once every 10$^9$ year. However, if consider a composite system of $N$ particles then Eq.(\ref{EQ:11}) is replaced by

\begin{equation}
\label{EQ:13}
\frac{d\rho}{dt} = - \frac{i}{\hbar} \left[ \hat{H}, \rho \right] + \lambda \sum_{i = 1}^N \left(T_i(\rho) - \rho \right),
\end{equation}

where $T_i(\rho)$ denotes the Gaussian weighing for the $i$-th particle localization operator $\mathbf{\hat{q}}_i$. If we define a collective variable, such as $\mathbf{Q}$ satisfying $\mathbf{q}_i = \mathbf{Q} + \sum_{j = 1}^{N-1} c_{ij} \mathbf{r}_j$, then the evolution of the system can be described using the ''reduced'' form $\rho_Q$ of the density matrix, defined by taking the partial trace over the relative localization variables $\mathbf{r}_j$,

\begin{align}
\label{EQ:14}
\rho_Q &= \text{tr}^{(\mathbf{r})}(\rho), \\
\nonumber
\rho_Q \left(\mathbf{Q}_1,\mathbf{Q}_2 \right) &= \int \prod d\mathbf{r}_k \cdot \rho( \mathbf{Q}_1 + \sum c_{ij} \mathbf{r}_j,\mathbf{Q}_2 + \sum c_{ij} \mathbf{r}_j).
\end{align} 

We observe that $\text{tr}^{(\mathbf{r})}(T_i(\rho)) = T_Q(\text{tr}^{(\mathbf{r})}(\rho))$ since,

\begin{align*}
\text{tr}^{(\mathbf{r})}(T_i(\rho)) = \\
\frac{1}{a^3 \pi^{3/2}}\int \int \prod d\mathbf{r}_k d\mathbf{x} \cdot 
e^{- \left( \mathbf{Q}_1 + \sum c_{ij} \mathbf{r}_j - \mathbf{x} \right)^2/{2a^2}} \\
\rho( \mathbf{Q}_1 + \sum c_{ij} \mathbf{r}_j,\mathbf{Q}_2 + \sum c_{ij} \mathbf{r}_j)e^{- \left( \mathbf{Q}_2 + \sum c_{ij} \mathbf{r}_j - \mathbf{x} \right)^2/{2a^2}} = \\
\frac{1}{a^3 \pi^{3/2}} \int d\mathbf{y} e^{- \left( \mathbf{Q}_1 - \mathbf{y} \right)^2/{2a^2}} \int \prod d\mathbf{r}_k  \\
 \rho( \mathbf{Q}_1 + \sum c_{ij} \mathbf{r}_j,\mathbf{Q}_2 + \sum c_{ij} \mathbf{r}_j)e^{- \left( \mathbf{Q}_2 - \mathbf{y} \right)^2/{2a^2}} = \\
T_Q(\rho_Q).
\end{align*}

Applying this to Eq.(\ref{EQ:13}) gives the modified SE for $\rho_Q$,

\begin{equation}
\label{EQ:15}
\frac{d \rho_Q}{dt} = - \frac{i}{\hbar} \left[ \hat{H}_Q,\rho_Q \right] + N \lambda \left( T(\rho_Q ) - \rho_Q \right).
\end{equation}

Here $\hat{H}_Q$ denotes the center-of-mass (CM) part of the Hamiltonian. For macroscopic processes ($N$ of the order of 10$^{23}$) we obtain a reduction process with the characteristic time $\tau = 1/N\lambda \approx 10^{-7}$s. A wave-packet $\Psi(\mathbf{Q})$, which describes (it's not clear what this ''describes'' exactly means in the GWR-approach) a macroscopic body represented by its CM-position $\mathbf{Q}$, ''contracts'' every 10$^{-7}$:th second to a region of the size $a \approx$ 10$^{-7}$m. In this model macroscopic bodies seem to have definitive CM-positions modulo circa 10$^{-7}$m; thus, a pointer wave function cannot exist in a superposition of two positions $\mathbf{Q}_1$ and $\mathbf{Q}_2$ satisfying $|\mathbf{Q}_1 - \mathbf{Q}_2 |  \gg 10^{-7}$m.
This conclusion holds only in the sense that the off-diagonal terms $\rho_Q \left(\mathbf{Q}_1, \mathbf{Q}_2\right)$ decreases exponentially; that is, the absence of superposition of the $\mathbf{Q}_1$- and the $\mathbf{Q}_2$-state takes an infinite time to reached. According to the definitions above we get,

\begin{align*}
\text{tr}^{(\mathbf{r})}(T(\rho)) = \frac{1}{a^3 \pi^{3/2}} \int d\mathbf{z} e^{- \mathbf{z}^2/{2a^2}}e^{- \left( \mathbf{Q}_2 - \mathbf{Q}_2 + \mathbf{z} \right)^2/{2a^2}} 
\rho_Q ( \mathbf{Q}_1,\mathbf{Q}_2) = \\
e^{- \left( \mathbf{Q}_2 - \mathbf{Q}_2\right)^2/{4a^2}} \rho_Q ( \mathbf{Q}_1,\mathbf{Q}_2),
\end{align*}

which in combination with Eq.(\ref{EQ:15}) results in

\begin{equation}
\label{EQ:16}
\rho_Q ( \mathbf{Q}_1,\mathbf{Q}_2) \approx \exp \left[ - N \lambda \left(1 - \exp \left(-|\mathbf{Q}_2 - \mathbf{Q}_2 |^2 /4a^2 \right) \right)t \right].
\end{equation}

This asymptotic decline of the interference does not satisfy Shimony. Besides the energy $E(t) = \text{tr}(\hat{H} \rho(t))$ is not conserved in the GWR-theory but increases with ca 10$^{-21}$J per second for a macroscopic system. This follows because a contracted wave packet represents a higher kinetic energy than a plane wave having the same momentum. The wave function $\phi$ for a free particle (plane wave) contracts $t/1/\lambda$ times during the time interval $t$ on a wave packet $\Psi$ of the dimension $a$. This is associated with a momentum dispersion of the size $\Delta p \approx \hbar/a$ corresponding to a kinetic energy ${\Delta p}^2 /2m \approx \hbar^2/2ma^2$:

\begin{align*}
E_\Psi = \left\langle \Psi \right| \frac{\hat{p}^2}{2m} \left| \Psi \right\rangle =\\
\left\langle \Psi \right| \frac{{\hat{p} - \left\langle p \right\rangle}^2}{2m} \left| \Psi \right\rangle + \frac{\left\langle p \right\rangle^2}{2m} \approx \\
\frac{\hbar^2}{2ma^2} + E_\phi,
\end{align*}

since $\left\langle \hat{p} \right\rangle_\Psi$ = $\left\langle \hat{p} \right\rangle_\phi$ and $\left\langle \hat{p}^2 \right\rangle_\phi$ = $\left\langle \hat{p} \right\rangle_\phi^2$ for a plane wave. The increase of energy during the interval $t$ thus becomes,

\begin{equation}
\label{EQ:17}
\Delta E = (E_\Psi - E_\phi) \frac{t}{1/\lambda} = \frac{\hbar^2 \lambda}{2ma^2}t.
\end{equation}

Apparently we need an energy input in order to sustain the GWR-process.

With regards to the modifications of the SE Shimony proposes at present only the general equation VACUUM = LETHE (= the river of oblivion), and suggests that the experimental results of the quantum telegraphy involving 3-level systems may provide constraints on the possible modifications of the SE. (The quantum telegraph is connected with the so called quantum Zeno effect \citep{Chiu1977,Block1991,Frerichs1991} which in the 3-level case has been treated as a system coupled to a reservoir. The Zeno effect denotes the situation where a continuous observation of an unstable system prevents it from decaying; the observation reduces the wave packet constantly to the initial state. The watched pot effect is another illustrative name for the effect.) 

The non-conservation of the energy (still quite minimal corresponding to a temperature increase of 10$^{-15}$K year for an ideal gas \citep{Ghirardi1986}) in the GWR-theory may indicate that it describes an open system and that the GWR-equation Eq.(\ref{EQ:11}) represents an ansatz to an ''effective'' SE for a system coupled to its environment. One might think of an analogy with the ''effective'' Lagrange-functions (or actions) used in QFT which incorporate some of the effects of vacuum polarizations and other interactions. From this point of view one ought to derive the ''effective'' SE from a fundamental theory of interacting particles, given that the ordinary SE is valid for closed systems and free particles. One aspect of this problem is studied for objects interacting with a reservoir \citep{Meystre1990}. All the decay processes can be related to this case. A classical model is formed by an harmonic oscillator $a$ (the object) coupled to a surrounding represented by a set of oscillators $b_k$, according to the Hamiltonian,

\begin{equation}
\label{EQ:18}
\hat{H} = \hbar \omega \hat{a}^{\dagger} \hat{a} + \sum_k \hbar \omega_k \hat{b}^{\dagger}_k \hat{b}_k + \sum_k \left\{ \lambda_k \hat{a}^{\dagger} \hat{b}_k + \lambda_k^{\star} \hat{a}\hat{b}^{\dagger}_k \right\}.
\end{equation}

Here $\hat{a}^{\dagger}$ denotes the creation operator of the $a$-oscillator, while $b$ refers to the oscillators representing the environment. The commutation relations are given by,

\begin{equation}
\label{EQ:19}
\left[\hat{a}, \hat{a}^{\dagger} \right] = 1, \quad 	\left[\hat{b}_k, \hat{b}_k^{\dagger} \right] = 1,
\end{equation}

the other pairs commuting.

If the above relations are inserted into the SE then we obtain,

\begin{align}
\label{EQ:20}
\frac{d \hat{a}}{dt} &= \frac{i}{\hbar} \left[\hat{H},\hat{a} \right] = -i\omega \hat{a} - \sum_k \frac{\lambda_k}{\hbar} \hat{b}_k , \\
\nonumber
\frac{d \hat{b}_k}{dt} &= \frac{i}{\hbar} \left[\hat{H},\hat{b}_k \right] = -i\omega \hat{b}_k - \frac{\lambda_k^{\star}}{\hbar} \hat{a} .
\end{align} 

Similar equations are obtained e.g. for the decay $K \rightarrow \pi + \pi$ \citep[ch.7]{Lipkin1973}, atomic transitions $A^{\star} \rightarrow A + \gamma$ (the Weisskopf-Wigner theory of spontaneous emission \citep{Weisskopf1930}). Generally Eq.(\ref{EQ:20}) quantum optical phenomena where a system (an atom) is coupled to a broad band system (reservoir) which absorbs the excitations and causes back reactions. Eq.(\ref{EQ:20}) is the foundation of the Weisskopf-Wigner equations, Kramers-Kronig relations, Pauli master-equations, quantum mechanical Langevin- and Fokker-Planck equations. The common property for these systems is that they represent irreversible decays (dissipation, friction, diffusion) whose time-asymmetric forms contradict the timesymmetric nature of the fundamental equations. The exponential decay cannot be derived from the SE in senso stricto, which has i.a. been observed by Eugen Merzbacher (quoted by \cite{Cartwright1983}): ''The fact remains that the exponential decay law, for which we have so much empirical support in radioactive processes, is not a rigorous consequence of quantum mechanics but the result of somewhat delicate approximations'' \cite[p.484-5]{Merzbacher1970}. Similar considerations apply in classical mechanics -- reversible Hamiltonian equations cannot lead to dissipative processes. An apparently elementary phenomenon such as friction seems to lack theoretical foundation. \citet{Cartwright1983} has employed examples like the above ones to argue for the primary role of the approximations (phenomenological equations) in physics at the expense of the exact (fundamental) equations, which seldom describe the empirical reality but rather some finely tuned models. Cartwright points out that \citet{Weisskopf1930} in the beginning only derived an exponential decay term from Eq.(\ref{EQ:20}). However, after the discovery 1947 by Willis Lamb and R C Retherford of the ''Lamb-shift'' between the levels $2^2 P_{1/2}$ and $2^2 S_{1/2}$ in hydrogen ($\Delta E$ = 0.437 10$^{-5}$eV = 1058 MHz) did Hans Bethe also find the Lamb-shift hidden in Eq.(\ref{EQ:20}) using a new approximation procedure \citep{Bethe1947}. Cartwright's conclusion is that ''the Schr\"odinger equation does not make a claim whether there is a Lamb shift in the circumstances described''. In a sense we use the equations in order to obtain results we already know, not in order to predict them; that is, we can calculate the transition probabilities using Born approximations etc, but the very fact of the transition as an event cannot be accounted for by quantum mechanics (which Einstein already pointed out). The GWR-model however introduces the reduction already on the micro-physical level. There are also models that introduce dissipation on the micro-physical level through non-linear modifications of the SE; one example is the ansatz (setting $\hbar$ = 1) 

\begin{equation}
\label{EQ:21}
\frac{d \Psi(t)}{dt} = -i \hat{H} \Psi(t) + k\left\{ \left\langle \Psi(t) |\hat{H} |\Psi(t) \right\rangle - \hat{H} \right\}\Psi(t),
\end{equation} 

by \citet{Gisin1990} where $k > 0$ is a quantum friction coefficient (for a generalization see \cite{Huang1989}). The problem with these models seems to be that the conventional theory approximates quite well the experimental results whence there is little room for modifications of the SE. Thus it has been suggested \citep{Gisin1990} that a non-linear SE \citep{Weinberg1989} may imply the possibility of superluminal commincations in contradiction with the special theory of relativity. \citet{Polchinski1991} argues that such modifications may also lead to communications (Everett phone) between different branches of a multi-verse wave functions. (His example involves a non-linear Hamilton operator which depends on the observer's choices, whence it is no surprise that he concludes that ''the apparatus reads the observer's mind''. In physics we had rather expect the ''observer'' to be described by an Hamiltonian.) As a conclusion we might say that the most reasonable interpretation of the non-linear modifications of the SE which implies energy non-conservation is that they are about open systems; the equations should therefore be based on some fundamental theory of interaction. 

One approach to the problems involving object-reservoir/environment-interaction, which seems to be shared by i.a. \citet{Peres1986} and \citet{Zurek1989}, is to be content with the approximative irreversible solutions of Eq.(\ref{EQ:20}). The idea is that we observe irreversible processes because we cannot observe the whole (reversible) process; information is lost to the environment (dissipation = displacement of information). The approximative assumptions leading from Eq.(\ref{EQ:20}) to the irreversible solutions are based on an assumption of some sort of memory loss; e.g., restricted correlation with the surrounding (Markov ansatz). In the classical analysis L Boltzmann introduced 1872 the Stosszahlenansatz according to which the colliding atoms (hard spheres in the model) loose their ''memory'' of their state before the collision. This entails an obvious time-asymmetry (for a simplified version of the Boltzmann analysis see \citep{Baker1986}). The classical Navier-Stokes equation in hydrodynamics is also based on a form of memory loss represented by the viscosity term (for derivations of such phenomenological equations from fundamental principles see \citep{Emch1986}). 
Theories and interpretations which attempt only to account for the practically measurable (in a statistical description) by using various forms of ''memory loss'' and coarse graining can be called FAPP theories (FAPP = For All Practical Purposes). For instance \citet{Zurek1982} is content with demonstrating the wave collapse in a FAPP theory.

Returning to Eq.(\ref{EQ:20}) we may solve for $b$ and insert the expression into the $a$-equation obtaining (setting $\hbar$ = 1) an operator equation,

\begin{equation}
\label{EQ:22}
\frac{d \hat{a}(t)}{dt} = -i \omega \hat{a}(t) - \int_0^t \phi(t - s) \hat{a}(s) ds + \hat{\epsilon}(t),
\end{equation}

with

\begin{align}
\label{EQ:23}
\phi(u) &= \sum_{k = 1}^N |\lambda_k|^2 e^{-i \omega_k u}, \\
\nonumber
\hat{\epsilon}(t) &= -i \sum_{k = 1}^N \lambda_k \hat{b}_k (0)e^{-i \omega_k u} .	
\end{align}
 
Eq.(\ref{EQ:22}) is called a quantum Langevin equation in analogy to the classical Langevin equation

\begin{equation}
\label{EQ:24}
m \frac{d \mathbf{v}}{dt} = -\beta \mathbf{v} + \mathbf{\eta}(t). 
\end{equation}

This equation contains a frictional term $-\beta \mathbf{v}$ and a stochastic background field $\mathbf{\eta}(t)$ (force) which simulates the influence of the environment. By introducing coherent states $\left| \alpha \right\rangle$ and $\left| \beta_k \right\rangle$ defined for complex numbers $\alpha$ and $\beta_k$ by

\begin{equation}
\label{EQ:25}
\hat{a} \left| \alpha \right\rangle = \alpha \left| \alpha \right\rangle , \quad \hat{b}_k \left| \beta_k \right\rangle = \beta_k \left| \beta_k \right\rangle ,
\end{equation}

we can write the operator equation as an ordinary integral equation. Using the orthonormal quantum states

\begin{equation}
\label{EQ:26}
\left| n \right\rangle = \frac{\left({\hat{a}}^\dagger \right)^n}{\sqrt{n!}} \left| 0 \right\rangle , \quad \left| n_k \right\rangle = \frac{\left({\hat{b}_k}^\dagger\right)^{n_k}}{\sqrt{{n_k}!}} \left| 0 \right\rangle ,
\end{equation}

we obtain

\begin{align}
\label{EQ:27}
\left| \alpha \right\rangle = \sum_n \left| n \right\rangle \left\langle n |  \alpha \right\rangle = \sum_n \frac{\left({\hat{a}}^\dagger\right)^n}{\sqrt{n!}} \left| 0 \right\rangle \left\langle 0 \right| \frac{{\hat{a}}^n}{\sqrt{n!}} \left| \alpha \right\rangle = \\
\nonumber
\left\langle 0 | \alpha \right\rangle \sum_n \frac{\left(\alpha {\hat{a}}^\dagger \right)^n}{n!} \left| 0 \right\rangle = \left\langle 0 | \alpha \right\rangle e^{\alpha {\hat{a}}^\dagger} \left| 0 \right\rangle.  
\end{align}

The normalization constant $c = | \left\langle 0 | \alpha \right\rangle |$ is determined by

\begin{align*}
1 = \left\langle \alpha | \alpha \right\rangle = c^2 \sum_{m,n} \left\langle n \right| \frac{{\bar{\alpha}}^n}{\sqrt{n!}} \frac{{\alpha}^m}{\sqrt{m!}} \left| m \right\rangle = c^2 e^{|\alpha|^2} ; \\
\text{that is,} \quad c = e^{-|\alpha|^2 /2} .
\end{align*}

Choosing the initial state

\begin{align}
\label{EQ:28}
\left| \alpha , \beta \right\rangle &= \left| \alpha \right\rangle \otimes \prod_k \left| \beta_k \right\rangle \\
\nonumber
&= \exp\left\{ \alpha {\hat{a}}^\dagger + \sum_k \beta_k {\hat{b}_k}^\dagger - |\alpha|^2 /2 - \sum_k |\beta_k|^2 /2    \right\} \left| 0 \right\rangle ,
\end{align}  

and defining the functions

\begin{align}
\label{EQ:29}
\alpha(t) &\equiv \left\langle \alpha , \beta \right| \hat{a}(t) \left| \alpha , \beta \right\rangle = \left\langle \alpha , \beta , t \right| \hat{a}(0) \left| \alpha , \beta , t \right\rangle ,\\
\nonumber
\beta_k(t) &\equiv \left\langle \alpha , \beta \right| \hat{\beta}_k(t) \left| \alpha , \beta \right\rangle = \left\langle \alpha , \beta , t \right| \hat{\beta}_k(0) \left| \alpha , \beta , t \right\rangle ,
\end{align}

then Eq.(\ref{EQ:20}) can be rendered on the form ($\hbar = 1$)

\begin{align}
\label{EQ:30}
\frac{d\alpha(t)}{dt} &= - i\omega\alpha(t) - i \sum_k \lambda_k \beta_k (t) ,\\
\nonumber
\frac{d\beta_k (t)}{dt} &= - i\omega_k \beta_k(t) - i \sum_k \lambda_k^\star \alpha(t) .
\end{align}

Finally, solving for $\beta_k$ and inserting the solution into the $\alpha$-equation, and defining $\gamma(t)$ by $\alpha(t) = \gamma(t) e^{-i \omega t}$ (the interaction picture), we end up with the integral equation

\begin{align}
\label{EQ:31}
\frac{d\gamma (t)}{dt} = - \int_0^t \sum_k |\lambda_k|^2 e^{-i(\omega_k - \omega)(t-s)} \gamma(s) ds \\
\nonumber
- i \sum_k \lambda_k \beta_k (0) e^{-i(\omega_k - \omega)t}.
\end{align}

Assuming as an initial condition that $\beta_k (0) = 0$ the second term can be neglected. The Lamb-shift and the exponential decay can be derived using the following assumptions:

\begin{enumerate}
\item the discrete frequency modes are replaced by a continuum making the sum into an integral $\int_{-\infty}^{\infty} |\lambda(\nu)|^2 \mathcal{D}(\nu) \dots \exp(-i(\nu \dots))$ where $\mathcal{D}(\nu)$ denotes the density of the $\nu$-states
\item we assume that $\gamma(t)$ varies slowly with time (weak coupling $\lambda$) and we replace $\gamma(s)$ with $\gamma(t)$ which is placed outside the integral
\item we let the integration interval go to infinity ($\int_0^t \rightarrow \int_0^\infty$) justifying it by the assumption that the greatest contribution comes from a small interval (Markov ansatz); that is, if $\lambda(\nu)$ is assumed to vary slowly then the integration over $\nu$ approximates a delta function $\delta (t - s)$.  
\end{enumerate}

Using these approximations we obtain

\begin{align}
\label{EQ:32}
\frac{d\gamma (t)}{dt} &= -\gamma(t) \int_{0}^{\infty} \int_{-\infty}^{\infty} \mathcal{D}(\nu) |\lambda(\nu)|^2 e^{-i(\nu - \omega)t}e^{i(\nu - \omega)s} d\nu ds\\
\nonumber
&= -\gamma(t) \int_{-\infty}^{\infty}  \mathcal{D}(\nu) |\lambda(\nu)|^2 e^{-i(\nu - \omega)t}\left\{\pi\delta(\nu - \omega) + i \mathcal{P}\left(\frac{1}{\nu -\omega} \right)  \right\} d\nu \\
\nonumber
&= -\gamma(t) \left\{\frac{\Gamma}{2} + i \Delta \omega \right\},   
\end{align}

where the decay rate is given by

\begin{equation}
\label{EQ:33}
\Gamma = 2 \pi \mathcal{D(\omega)} |\lambda(\omega)|^2 \quad \text{(''Fermi golden rule'')}
\end{equation}

and the Lamb-shift by

\begin{equation}
\label{EQ:34}
\Delta \omega = - \int_{-\infty}^{\infty} \frac{|\lambda(\omega)|^2 \mathcal{D}(\nu) d\nu}{\omega - \nu}
\end{equation}

(supposing we can neglect the factor $\exp(-i(\nu - \omega))$ since the integrand is weighed around $\nu \sim \omega$). The approximative and asymptotic solution for the function $\alpha(t)$ may thus be written

\begin{equation}
\label{EQ:35}
\alpha(t) \approx \alpha(0) \cdot e^{-i(\omega + \Delta \omega)t - \Gamma t /2} .
\end{equation}

The above model does not indicate that a system in contact with a reservoir necessarily behaves in a ''classical'' manner. The GWR-model in contrast suggests that a macroscopic body of mass, say 1 kg, will closely follow a  classical path with velocity fluctuations of the order of $\delta v \sim \hbar/1 \text{kg} 10^{-7} \text{m} \sim 10^{-27}$m/s. 

\epigraph{I do not know what the Copenhagen interpretation is but I know Copenhagen -- it's a very beautiful city}{A Peres, at the conference}
\section{Decoherence}

\citet{Omnes1990} has used the above reservoir model in order to demonstrate how the influence of the environment may suppress the superposition states of the object. Consider two different initial states $|q_1 - q_2| \gg 0$ of the object (oscillator), representing the average values

\begin{align}
\label{EQ:36}
q_i &= \left\langle \alpha_i \right| \hat{q} \left| \alpha_i \right\rangle = 
\left\langle \alpha_i \right| \sqrt{\frac{\hbar}{2 M \omega}} \left( \hat{a} + {\hat{a}}^\dagger \right) \left| \alpha_i \right\rangle \\
\nonumber
&=  \sqrt{\frac{\hbar}{2 M \omega}} \left( \alpha_i + \alpha_i^\star \right) 
\quad (i = 1, 2) ,
\end{align}

for the position operator $\hat{q}$. According to quantum mechanics the system may exist in a superposition state

\begin{equation}
\label{EQ:37}
\left| \Psi \right\rangle_{t = 0} = \frac{1}{\sqrt{2}} \left\{ \left| \alpha_1 \right\rangle + \left| \alpha_2 \right\rangle \right\} \otimes \left| \text{vacuum} \, (\beta = 0) \right\rangle .
\end{equation}

We can form the reduced density matrix $\rho$ for the oscillator $a$ by taking the trace of $\rho = \left| \Psi \right\rangle \left\langle \Psi \right|$
over the $b$-oscillators,

\begin{equation}
\label{EQ:38}
	\tilde{\rho} = \text{tr}^{(\beta)} \left( \left| \Psi \right\rangle \left\langle \Psi \right| \right) =
	\int \left( \prod_k \frac{d^2 \beta_k}{\pi} \right) \left\langle \beta | \Psi \right\rangle \left\langle \Psi | \beta \right\rangle .
\end{equation}

For asymptotic times we may assume that the initial coherent state $\left| \alpha , \beta \right\rangle$ evolves on $\left| \alpha(t) , \beta(t) \right\rangle$ where $\alpha(t)$ and $\beta_k(t)$ are determined by Eq.(\ref{EQ:30}) and the approximative solution Eq.(\ref{EQ:35}). This yields,

\begin{equation}
	\label{EQ:39}
	\tilde{\rho} \sim \frac{1}{2} \sum_{i,j} \left\langle \beta_{(i)} | \beta_{(j)} \right\rangle_t \left| \alpha_j e^{-i \acute{\omega}t - \Gamma t/2} \right\rangle \left\langle \alpha_i e^{-i \acute{\omega}t - \Gamma t/2} \right| .
\end{equation}

From the definitions it follows that $| \left\langle \beta_{(i)} | \beta_{(j)} \right\rangle |$ is equal to

\begin{equation*}
\exp\left\{ - \frac{1}{2} \sum_k | \beta_{(i)} - \beta_{(j)} |^2 \right\}.
\end{equation*}

We observe that $\alpha_{12}(t) \equiv = \alpha_1(t) - \alpha_1(t)$ and
$\beta_{k,12}(t) \equiv = \beta_{1(k)}(t) - \beta_{2(k)}(t)$ again satisfy the Eq.(\ref{EQ:30}) with the integral

\begin{equation*}
|\alpha_{12}|^2 + \sum_k |\beta_{12(k)}|^2 = \text{const.} = |\alpha_{12}(0)|^2 .
\end{equation*}

From this it follows that

\begin{equation*}
\sum_k |\beta_{1(k)} - \beta_{2(k)}|^2 \approx |\alpha_1(0) - \alpha_2(0)|^2 \cdot (1 - e^{- \Gamma t}) .
\end{equation*}

Assuming the initial values $\alpha_i(0)$ are real quantities we obtain finally

\begin{equation}
\label{EQ:40}
| \left\langle \beta_{(i)}(t) | \beta_{(j)}(t) \right\rangle | \approx 
\exp\left\{ - \frac{M \omega}{4 \hbar} |q_1 - q_2 |^2 (1 - e^{-\Gamma t}) \right\}.
\end{equation}

Omn\`es concludes that the superposition of states 1 and 2 ($|q_1 - q_2 | \gg 0$) is suppressed for large enough time $t > 0$. 
 
Common to all the decoherence models, the above included, is that they cannot prove the superposition between macroscopic states vanishes in a finite time. (Several models employ the Riemann-Lebesgue lemma in order to show that $\rho_{\text{off-diag}} \rightarrow 0$ when $t \rightarrow \infty$, vide \citep{Machida1980}). Dieks accepts the dechoerence models as solutions to the measurement problem in union with his ontological assumption; the decoherence means that the ''real/physical'' states approaches definitive values in the measurement process. Omn\`es observes that the exact solutions to Eq.(\ref{EQ:30}) (for a finite number $N$ of oscillators) predict a Poincar\'e recurrence; that is, for any given time $t > 0$ there is a time $T > t$ such that ($\alpha(T), \beta(T)$) approaches the initial state ($\alpha(0), \beta(0)$) arbitrarily closely which contradicts the exponential law of decay Eq.(\ref{EQ:35}). (For atoms which interacts with single mode fields, $N$ = 1, one obtains explicitly periodic solutions which manifested in Rabi-flopping, or Jaynes-revival of apparently collapsed atomic states, see \cite{Meystre1990}). According to Omn\`es this means that even classical physics entails a fuzzy logic (since Poincar\'e recurrence applies to classical mechanics, see \cite{Arnold1978}) which are in harmony with the ''apparent'' validity of the decay law. He compares the statistical probability of the Poincar\'e recurrence with the probability that the Earth would suddenly move, through quantum tunneling, and start to circle around Sirius. 

A general class of decoherence models attempts to demonstrate that the superposition between two, say orthogonal, quantum states $\left| \phi_1 \right\rangle$ and $\left| \phi_2 \right\rangle$ of an object is suppressed by  a coupling to the environment ($\left| \eta \right\rangle$), by which the superposition is turned into

\begin{equation}
\label{EQ:41}
\left| \Psi \right\rangle = \frac{1}{\sqrt{2}} \left\{ \left| \phi_1 \right\rangle  \otimes \left| \eta_1 \right\rangle + \left| \phi_2 \right\rangle \otimes \left| \eta_2 \right\rangle \right\} .
\end{equation}
 
The idea is to show that $\left\langle \eta_1 | \eta_2 \right\rangle \sim 0$ (compare with $\left\langle \beta_{(i)} | \beta_{(j)} \right\rangle$ in Eq.(\ref{EQ:40})) which is equivalent to the reduced density matrix

\begin{align}
\label{EQ:42}
\tilde{\rho} &\equiv \text{tr}^{(\eta)} \left( \left| \Psi \right\rangle \left\langle \Psi \right| \right)\\ 
\nonumber 
&= \frac{1}{\sqrt{2}} \left\{
\left| \phi_1 \right\rangle \left\langle \phi_1 \right| +
\left| \phi_2 \right\rangle \left\langle \phi_2 \right| +
\left| \phi_1 \right\rangle \left\langle \phi_2 \right|
\left\langle \eta_2 | \eta_1  \right\rangle +
\left| \phi_2 \right\rangle \left\langle \phi_1 \right|
\left\langle \eta_1 | \eta_2  \right\rangle
\right\}
\end{align}

approaching the diagonal form

\begin{equation}
\label{EQ:43}
\tilde{\rho}_d =
\frac{1}{\sqrt{2}} \left\{
\left| \phi_1 \right\rangle \left\langle \phi_1 \right| +
\left| \phi_2 \right\rangle \left\langle \phi_2 \right| \right\} .
\end{equation}

This may be interpreted as a proof for the wave collapse; that is, that the superposition Eq.(\ref{EQ:41}) has collapsed on either of the states $\left| \phi_1 \right\rangle$ or $\left| \phi_2 \right\rangle$. The density matrix provides us though only with a statistical description (averages and probabilities) and cannot hence prove that an actual collapse takes place in which a superposition $\sum c_n \left| \phi_n \right\rangle$ ends up in a definitive state $\left| \phi_k \right\rangle$. It can thus not predict an event (the object assuming definitive states/values). A statistical description may allow a body to be at one place at one moment, and then at another moment -- with no continuous passage through the space -- to appear at another place (whose position is only statistically determined). This argument is presented i.a. by Cartwright as a demonstration of the need to accept the wave collapse as a physical process and not merely as a statistical description. \citet{Bell1987a} has moreover observed that it is always possible in principle to find an operator $\hat{O}$

\begin{equation}
\label{EQ:44}
\left\langle \phi_1 , \eta \right| \hat{O} \left| \phi_2 , \eta \right\rangle_{t = 0} \gg 0 ,
\end{equation}

by which one can construct $\hat{O}(t) \equiv e^{- i \hat{H}t/{\hbar}} \hat{O} e^{ i \hat{H}t/{\hbar}}$, where $\hat{H}$ is the Hamiltonian which describes the evolution of the object + environment during the measurement process. This operator $\hat{O}(t)$ satisfies

\begin{align}
\label{EQ:45}
\left\langle \phi_1 (t) , \eta(t) \right| \hat{O}(t) \left| \phi_2 (t) , \eta(t) \right\rangle 
&=
\left\langle \phi_1 , \eta \right| e^{ i \hat{H}t/{\hbar}} \hat{O}(t) e^{ -i \hat{H}t/{\hbar}} \left| \phi_2 , \eta \right\rangle \\
\nonumber
&=
\left\langle \phi_1 , \eta \right| \hat{O} \left| \phi_2 , \eta \right\rangle \gg 0 .
\end{align}
 
Thus by measuring the operator $\hat{O}(t)$ we can detect an interference between ($\phi_1, \eta_1$) and ($\phi_2, \eta_2$); that is, the measurement of  
$\hat{O}(t)$ will demonstrate the difference between the superposition state $\rho = \left| \Psi \right\rangle \left\langle \Psi \right\rangle$ and the reduced form. We get

\begin{equation*}
\text{tr}\left({\hat{O} \rho(t)} \right) = 
\left\langle \Psi \right| \hat{O}(t) \left| \Psi \right\rangle \sim
\frac{1}{2} \sum_{i,j}
\left\langle \phi_i , \eta_i , t \right| \hat{O}(t) \left| \phi_j , \eta_j , t \right\rangle ,
\end{equation*}

while the reduced form $\tilde{\rho}$ Eq.(\ref{EQ:42}) neglects all observables associated with the environment. Supporters of the FAPP-models, such as Omn\`es, rejoin that the proposed observables $\hat{O}(t)$ are in practice not measurable. The probability that such ''supermeasurements'' will yield non-trivial results is ''much smaller than the limit one must accept for classical logic'' \citep{Omnes1990}. Asher Peres agrees -- it is impossible for us to keep a tab on all the environmental variables. The wave collapse, according to Peres, is nothing more than the result of the necessity to restrict the observation to a subsystem in which the loss of information will be manifested by reduced density matrices and equations involving irreversibility. For Peres the interpretational problem in quantum mechanics concerns a translation from the language of Hilbert space structure to the language of the classical phase space. This translation from the language of quantum mechanics to the language of classical physics then gives rise to the irreversibility as an aspect of the languages (displacement of information). Quantum mechanics with its unitary time evolutions is not a theory of events, according to Peres -- he defines quantum mechanics as ''a limiting case for infinite systems with no back-reactions''. Similarly \citet{Zurek1982} also links the wave packet reduction to a limited information capacity. One consequence of Zurek's ideas is that if 

\begin{equation}
\label{EQ:46}
\left\langle \Psi_1 \right| \hat{O} \left| \Psi_2 \right\rangle \sim 0
\end{equation}

is satisfied for two states $\left| \Psi_1 \right\rangle$, $\left| \Psi_2 \right\rangle$, for practically measurable observable $\hat{O}$ (for which one can construct a corresponding measurement device), then

\begin{equation}
\label{EQ:47}
\left| \Psi \right\rangle = \frac{1}{\sqrt{2}} \left\{ \left| \Psi_1 \right\rangle + \left| \Psi_2 \right\rangle \right\} 
\end{equation}

is not a proper superposition state. Instead its meaning is that the object either is in the state $\left| \Psi_1 \right\rangle$ or $\left| \Psi_2 \right\rangle$ (with 50 \% chance for either). The condition Eq.(\ref{EQ:46}) entails that $\left| \Psi_1 \right\rangle$ and $\left| \Psi_2 \right\rangle$ are ''effectively superselected'', and it can be compared with the Principle of State Distinction (PSD) advanced by Schr\"odinger: states of macroscopic systems that can be told apart by macroscopic observations remain separated whether observed or not \citep[p.216]{Jammer1974}; that is, the state Eq.(\ref{EQ:47}) cannot be a superposition state. This criterion can be applied to the Schr\"odinger cat state

\begin{equation}
\label{EQ:48}
\left| \Psi \right\rangle = \frac{1}{\sqrt{2}} \left\{ \left| \text{dead} \right\rangle + \left| \text{living} \right\rangle \right\} ,  
\end{equation}

when it is assumed that the states ''dead'' and ''living'' can be distinguished through a non-obtrusive observation. The PSD-principle might be connected to the superselection condition in the following way: PSD is based on the assumption that the observation of the macroscopic system changes it only by adding a phase factor and that this does not change the physical situation. According to this, the states

\begin{equation}
\label{EQ:49}
\left| \Psi \right\rangle = \frac{1}{\sqrt{2}} \left\{ \left| \Psi_1 \right\rangle + \left| \Psi_2 \right\rangle \right\} \quad \text{and} \quad
\left| \Psi \right\rangle = \frac{1}{\sqrt{2}} \left\{ \left| \Psi_1 \right\rangle + e^{i \delta}\left| \Psi_2 \right\rangle \right\} 
\end{equation}

should be equivalent for real numbers $\delta$. This means that no interference can be detected between $\left| \Psi_1 \right\rangle$ and $\left| \Psi_2 \right\rangle$; that is, the superselection condition Eq.(\ref{EQ:46}) is valid for every measurable operator. In the case of Eq.(\ref{EQ:48}) the condition $\left\langle \text{dead} \right| \hat{O} \left| \text{living} \right\rangle \sim 0$ implies that the cat cannot be revived using any such operator $\hat{O}$. The superselection condition in the sense of Zurek is linked
to irreversible transitions $\left| \Psi_1 \right\rangle \rightarrow \left| \Psi_2 \right\rangle$ where no realizable operator can return $\left| \Psi_2 \right\rangle$ to $\left| \Psi_1 \right\rangle$. The ''classical'' properties of systems/bodies are the ones superselected by the environment. This contextual definition of system properties has been elaborated within the $C^\star$-algebra approach to the quantum theory of ''infinite'' systems \citep{Primas1983}

\epigraph{I am not interested in the definition of science, but in the brains of the scientists}{E Huhmar, neurophysiologist, at the conference}
\section{Mind and physics}

Abner Shimony distinguishes two traditions within philosophy and science:

\begin{itemize}
	\item One tradition which tries to ''close the circle'' between epistemology and metaphysics; theories should not only describe the phenomena but also explain their existence and connect them with their ontological basis. Within this tradition Shimony counts members like Einstein, Aristotle, Leibniz and Whitehead.
	\item The other tradition studies epistemology (the condition for knowledge and phenomena) without an ontological grounding. As representative members Shimon mentions Bohr, Kant, Hume, Nietzsche and the linguistically oriented philosophers (Wittgenstein?).\footnote{Shimony's two traditions are not unlike the Galilean tradition (mechanistic, causal, predictions) and the Aristotelian tradition (teleological, understanding) that were analyzed by \citet{Wright1971}.} 
\end{itemize}

Shimony expresses sympathy for the former alternative since it is, from a ''global'' perspective, more fruitful, has a larger explanatory force and aims at a coherent world view. These arguments are not unlike the ones presented by Max Planck in support of realism (in \textit{Vortr\"age und Erinnerungen}, 1949, discussed in \citep{Howard1979}). Shimony presents a parable (a movie projected on a screen) which can be compared to the cave parable invented by Plato where chained prisoner in the cave were only able to see shadow of people on the walls. The movements of the shadows seem chaotic and incomprehensible till they realize that the shadows are produced by real bodies and their movements. According to Shimony it is equally reasonably to assume that all physical phenomena have an ontological grounding which can explain their cause and nature (a causal argument for micro-objects). By applying this philosophy to the wave collapse in quantum mechanics it follows that we should attempt at finding a physical explanation for the collapse, for instance by modifying the SE as suggested by Shimony.

Professor K V Laurikainen, who has played a central role in arranging the physics symposia in Finland, has followed a different track in his lectures. He thinks the question of reality is a psychological question. The meaning of the wave collapse is that the observer becomes aware of the measurement results; the wave collapse takes place in the mind of the observer, according to Laurikainen. From this point of view one may understand the claims that the psyche must be taken into consideration in discussing the foundations of quantum mechanics. Peres rejoins that consciousness is only involved in the presentation of the measurement data, not in the measurement process itself. Shimony is more sympathetic and he would prefer a mentalistic ontology (objective idealism) to pure physicalism, but he cannot accept Laurikainen's point of view which he summarizes as that the consciousness causes the wave collapse. However, Laurikainen does not accept this interpretation and emphasizes that the collapse takes place \textit{in the consciousness}.
Yet if consciousness has no causal role in the wave collapse then there are few reasons to believe that it plays any physical role whatsoever in the measuring process (besides the ''trivial'' fact that physicists and engineers have designed the experiments and the devices). The assumption that the wave collapse takes place in the consciousness implies in the end that everything takes place in the consciousness, all objectifications are a mental process. That leads us back to Machian positivism which defines physics as a part of psychology, as an economic description of perception data and mental states. Perhaps not such an inspiring idea for physics.

The measurement process involves fundamental physical principles which are responsible for the Pauli-exclusion (explaining e.g. electron ''shells''), superfluidity, radioactive decay, Bragg-interference, etc. In none of these cases is the mind presumed to be a causal factor. On the contrary, the physical theories have been developed to explain why we see all these different phenomena. However, by this it is not implicated that consciousness can be reduced to the motion of electrons and protons (or behaviouristic reflexes of the larynx). It seems obvious that thoughts/feelings are not equivalent to (but probably contingent/supervenient) with particle states in the way e.g. an electric current is equivalent with a collective motion of electrons. This is the lasting Cartesian insight (or ''common sense'') into the dualism between res extensa (matter) and res cogitans (soul). Quantum mechanics has not been able to transcend this dualism despite some quantum mystical trends \citep{Zukav1979,Talbot1988,Zohar1990,Restivo1983}. Yet it is natural to expect that quantum mechanics as a fundamental theory will have consequences for the understanding of the ''psycho-physical parallel''; that is, the ''correlation'' between the physical and mental states (the mind-body problem). Thus, \citet{Wigner1961} envisaged that the mind implicates a non-linear SE since it cannot be described in terms of the conventional linear SE. According to his friend-argument an observer cannot exist in a superposition of the form

\begin{equation}
\label{EQ:50}
\left| \Psi \right\rangle = \frac{1}{\sqrt{2}} 
\left\{
\left| \text{has observed} \, x \right\rangle +
\left| \text{has not observed} \, x \right\rangle
\right\} 
\end{equation}

which is presumed by the conventional quantum mechanical formalism in the Schr\"{o}dinger cat cases. Wigner introduces consciousness as a causal agent as it causes the wave collapse. Another approach is to attempt at constructing quantum mechanical models for the ''brain states''. Thus, \citet{Marshall1989} has suggested a model for brain states based on the boson condensate model presented by \citet{Frolich1968}. (Such approaches will be discussed in a separate paper. For a discussion of quanta, mind and brain see \citep{Lockwood1989,Penrose1989}.) Laurikainen however does not express any interest in such models. He sees the lack of strict causality in nature as an evidence of a pan-psyche in the nature. While Descartes and Newton found evidence of the hand of God in the lawfulness of the universe,\footnote{The Newtonian gravitation was seen as an evidence for an invisible spiritual reality and was employed as an argument against atheists \citep{Jacob1988}. With the entrance of quantum mechanics matter becomes spiritualized -- \citet{Zukav1979} suggests that quantum particles are conscious -- the cricket ball matter is just a myth \citep{Davies1991} whence the genius of atheism has been defeated $\dots$} Laurikainen finds clues for pantheism in the ''irrational'' quantum indeterminism. Without arguing against or for pan-psychism one may still be inclined to think that a direct philosophical-theological linking to quantum mechanics would put an end to physics (Laurikainen for instance rejects the application of quantum mechanics to cosmology by pointing to the constraining concept of the observer \citep[p.136]{Laurikainen1991}). However there are still jobs to do for theoretical physics; for instance, little attention has been paid to the problem of deriving ''classical'' collective variables from quantum mechanics. One possibility is that quantum mechanics is not valid for macroscopic bodies. This question has been studied in a series of papers by \citet{Leggett1986} in connection with quantum effects in superconducting materials. Finally, the question of the relation between quantum mechanics, relativity and gravitation still remains wide open. We expect to return to this question when the appropriate occasion arises. ---

\epigraph{I hate quotations}{R. W. Emerson} 
\putbib[kvantprof.bib]
\end{bibunit}

\begin{bibunit}[apalike]

\section{Postscript (2006)}  

\subsection{Afterthoughts} 
In the paper there are expressions of the form $x \gg 0$ that surely seems odd to a mathematician. Of course, in physics, what is ''big'' and ''small'' depends on the context. Generally it is a question of comparing the order of magnitudes. Thus, e.g. $x \sim 0$ may be understood as $\log(|x|) \ll -1$. What is ''small'' relates to the question on what can be considered as ''negligible'' in a given context. In physics we cannot expect absolute certainty or infinite accuracy. But where to draw the boundary between blissful ignorance and anomalies? If a theory or a model predicts quantities to a precision of $\epsilon$, then any experimental deviation larger than $\epsilon$ would naturally be interpreted as a refutation of the theory/model. Because of our finitude (one of the Kantian themes) there appears to be a limit to the precisions $\epsilon$ we can attain. For instance the energy non-conservation implied by the GWR-process seems to be negligible for all practical purposes. On the other hand, the discussions of fine tunings related to the cosmological ''constant'' $\Lambda$ have revealed that even the 100th decimal of $\Lambda$ may be of far reaching consequence for the evolution of the universe. As far as the physical world (or rather our description of it) is subjected to stringent laws, expressed e.g. in terms of differential equations, problems involving fine tuning can hardly be avoided. Of course it is a logical possibility that our ''laws of physics'' are valid only in some restricted range of parameters and that there is no bottoming out. While our \textit{descriptions} of nature in physics are formulated in terms of more or less rigorous and exacting mathematics \citep{Steiner1998}, nature is not mathematics (pardon Platon!). Mathematics and experiments give a measure of how ''close'', in a sense, our descriptions and the nature (or some local ''domain'' of it) touch each other on some point. Perhaps a successful ''theory of everything'' would just be another reminder of that we are enclosed in a ''bubble'' of ignorance, stranded on a ''selfcontained'' island, or chained prisoners in Platon's metaphorical cave?

Below we have added some new references since 1992 on the problems of quantum mechanics and the philosophy of science:  \citet{Amann1999,Cohen1999,Cohen2006,Duplantier2006,Elitzur2005,Espagnat2005,Espagnat2006,Omnes1999,Omnes2004,Peres1995,Peres2006,Shimony1993}.

\subsection{K V Laurikainen (1916--1997)}
Prof K V Laurikainen was a primus motor behind the series of symposia on the foundations of modern physics that were arranged in Finland 1977 -- 1994, and which attracted the topnotch investigators in quantum mechanics from over the world.\footnote{After the 1977 conference Pekka Lahti suggested to Laurikainen the idea of a conference devoted especially to the foundations of quantum mechanics to which Laurikainen responded enthusiastically. While Laurikainen pulled the strings, Lahti was mainly responsible for the scientific programmes for the 1985, 1987, and 1989 SFMP-conferences in Joensuu. Due to diverging scientific interests in the early 90's, Lahti, together with P Mittelstaedt and P Busch organized separate conferences in Cologne \citep{Busch1993}, while Laurikainen and Claus Montonen arranged symposia in Helsinki. In a sense the impetus for the SFMP-conferences can be traced back to Rolf Nevanlinna whose lectures 1946-7 on quantum mechanics had instilled Laurkainen with a permanent interest in the foundational issues of quantum mechanics. ''Personally I would say, that this lecture series [on QM], besides the introductory courses by Kalle Huhtala, was the most rewarding lectures in physics which I have ever experienced in Finland'' \citep{Laurikainen1982}.} 

As a student Laurikainen was greatly influenced by the mathematician Rolf Nevanlinna and the philosopher Eino Kaila \citep{Kajantie1997,Laurikainen1982} both who with a deep interest in theoretical physics. In 1947 Nevanlinna arranged for a scholarship enabling Laurikainen to study in Z\"urich where he for the first time met Wolfgang Pauli whose writings would later become one the chief sources for Laurikainen's philosophical inspiration. Under the guidance of prof Torsten Gustafson (Lund, Sweden) Laurikainen completed a thesis in 1950 on the gravitational energy of electromagnetic fields. Lamek Hulth\'en then got him interested in the problem of the deuteron which lead him to become an expert in computational physics. Back in Finland, first as a teacher and professor in Turku, then in Helsinki (1960-79), he became one of the main champions, in company with Nevanlinna and prof Karl-Gustav Fogel ({\AA}bo Akademi, Turku), of theoretical physics as a field within physics. At that time physics in Finland was predominantly experimental. In 1961 it became possible to major in theoretical physics. The Research Institute for Theoretical Physics started in 1964 and it got its own building in 1969 called the ''Laurikainen building''. Besides launching computational physics Laurikainen also played an instrumental role in getting Finland involved in particle physics and later to become a member of CERN. Thus many physicists have felt a great obligation toward Laurikainen for his contribution to the postwar reconstruction of physics research in Finland. When Laurikainen retired he embarked on equally vigorous projects centered on the study of Wolfgang Pauli's philosophical writings \citep{Laurikainen1988,Laurikainen1997,Ketvel1996}. In fact, he was scheduled to defend his thesis \citep{Laurikainen1997a} in philosophy on 22. August 1997, when he died a month before (13.7). While Laurikainen's philosophical project attracted little support among his colleagues and former students, Laurikainen was able to help arrange a successful series of international conferences in Finland on the foundations of physics. These symposia, centered on the foundational question of quantum mechanics, were right on the target at that time. Interpretational issues were again coming to the forefront of physics, one reason being the improved experimental methods for probing questions such as entanglement. It was also becoming possible to mention ''philosophy'' and ''metaphysics'' at physics conferences without being immediately labeled a crackpot. The symposia also succeeded in assembling an older generation of physicists who had had a direct interaction with the ''founders'' such as Heisenberg and Bohr. As a student I think I was quite fortunate in having this opportunity, via the conferences, to follow the discussions by many great physicists from around the world. I remember names (in no particular order) such as K Bleuler, C F v Weizs\"acker, Y Neeman, J P Vigier, D Bohm, A Zeilinger, H Stapp, R B Griffiths, S Kochen, C Piron, P Busch, P Mittelstaedt, P Lahti, M Jammer, N Rosen, A Aspect ... to name only a few. Again the legacy of Laurikainen was that he created a framework where people could meet and discuss new and old ideas and findings. Despite the fact that he hardly ever got a wholehearted support for his philosophical views, and was sometimes severely criticized, he remained polite and supportive toward everyone interested in discussing the issues. Laurikainen remained faithful to the ''Copenhagen interpretation'' but criticized Bohr for having avoided the ontological questions.

\putbib[kvantprof.bib]
\end{bibunit}

\end{document}